\documentclass[aps,prb,reprint,a4paper,floatfix,showpacs,showkeys,superscriptaddress]{revtex4-1}

\usepackage{graphicx}
\usepackage{amsmath}
\usepackage{epstopdf}
\usepackage{color}

\newcommand{\mytilde}{\raise.17ex\hbox{$\scriptstyle\mathtt{\sim}$}}

\begin{document}

\title{Efficient Generation of Model Bulk Heterojunction Morphologies for Organic Photovoltaic Device Modeling}

\author{Michael C. Heiber}
\email{michael.heiber@physik.uni-wuerzburg.de}
\affiliation{Department of Polymer Science, The University of Akron, Goodyear Polymer Center, Akron, Ohio 44325, USA}
\affiliation{Experimental Physics VI, Julius-Maximilians-University of W{\"u}rzburg, Am Hubland, D-97074 W{\"u}rzburg, Germany}

\author{Ali Dhinojwala}
\email{ali4@uakron.edu}
\affiliation{Department of Polymer Science, The University of Akron, Goodyear Polymer Center, Akron, Ohio 44325, USA}

\date{\today}

\begin{abstract}
Kinetic Monte Carlo (KMC) simulations have been previously used to model and understand a wide range of behaviors in bulk heterojunction (BHJ) organic photovoltaic devices, from fundamental mechanisms to full device performance.  One particularly unique and valuable aspect of this type of modeling technique is the ability to explicitly implement models for the bicontinuous nanostructured morphology present in these devices.  For this purpose, an Ising-based method for creating model BHJ morphologies has become prevalent.  However, this technique can be computationally expensive, and a detailed characterization of this method has not yet been published.  Here, we perform a thorough characterization of this method and describe how to efficiently generate controlled model BHJ morphologies.  We show how the interaction energy affects the tortuosity of the interconnected domains and the resulting charge transport behavior in KMC simulations.  We also demonstrate how to dramatically reduce calculation time by several orders of magnitude without detrimentally affecting the resulting morphologies.  In the end, we propose standard conditions for generating model morphologies and introduce a new open-source software tool.  These developments to the Ising method provide a strong foundation for future simulation and modeling of BHJ organic photovoltaic devices that will lead to a more detailed understanding of the important link between morphological features and device performance.
\end{abstract}



\maketitle

\section{Introduction}

Organic photovoltaics (OPVs) have received a great deal of attention over the last decade.  In this time, research efforts covering a wide range of challenges have pushed the power conversion efficiency of these devices from \mytilde3\% to \mytilde11\%.\cite{kazmerski2014}  Among these efforts, modeling and simulation have been important for testing our understanding of the fundamental physics of device operation and directing experimental efforts towards new and improved devices.  Within the wide range of methods available, kinetic Monte Carlo (KMC) simulations are unique in their ability to incorporate nanoscale details while maintaining the ability to simulate a complete device.  The complex nanoscale morphologies present in bulk heterojunction (BHJ) solar cells have been repeatedly shown to have a significant impact on device performance for a number of different donor-acceptor combinations, including polymer-fullerene blends,\cite{chen2009} polymer-polymer blends,\cite{mcneill2007} and small molecule blends.\cite{zheng2010}  As a result, retaining nanoscale detail in KMC simulations is particularly critical for incorporating morphological features that can be used to help understand how morphology affects device performance in greater detail.  

In an attempt to generate model morphologies for small molecule blends, \citeauthor{peumans2003}\cite{peumans2003} introduced a method that utilizes the Kawasaki spin-exchange Ising model,\cite{kawasaki1966a,*kawasaki1966b,*kawasaki1966c} which had previously been used to simulate phase separation in binary alloys.\cite{binder1974a,bortz1974} This concept was then later simplified and applied to KMC simulations by \citeauthor{watkins2005}\cite{watkins2005}  Since these pioneering studies, Ising-based morphologies have been used in KMC simulations to study a wide range of important OPV topics, from detailed studies on exciton diffusion and dissociation,\cite{mcneill2007,feron2012a,heiber2013c} charge separation and geminate recombination,\cite{groves2008a,groves2010a,groves2010b,yan2010} bimolecular recombination,\cite{groves2008b} surface recombination,\cite{feron2012b} and charge injection\cite{groves2009,kipp2013} to broader studies on overall photocurrent generation\cite{marsh2007,lei2008,yang2008} and complete current-voltage curve modeling.\cite{meng2010,kimber2010,meng2011,kimber2012,kipp2013}   Ising-based morphologies have also been used in master equation device modeling.\cite{koster2010}

Several additional morphology models have also been used in KMC simulations, including a chain reptation model\cite{frost2006,deibel2009} and a Cahn-Hilliard model.\cite{henderson2005,lyons2011,lyons2012}  In addition, the Ising model has been adapted to produce morphologies similar to those measured by neutron reflectivity and neutron scattering experiments.\cite{olds2012}  While the Ising model may not accurately capture all morphological features in all donor-acceptor blends used in OPVs, qualitatively, it produces a nanoscale bicontinuous morphology typical of many blends.  As a result, this model has served as a reasonable approximation of BHJ morphologies and has become the dominant morphology model in the field.  

Nonetheless, a rigorous characterization of this morphology generation technique has not been published.  Understanding the details of this method, developing standard procedures, and making the technique openly available will ensure that accurate benchmarks and meaningful comparisons are being made.  In the Ising model, the main parameter that can be used to control the phase separation process is the interaction energy.  However, it is still unclear exactly how changing this parameter impacts the morphologies generated and the resulting simulated device performance.  

In addition, making this technique more computationally efficient has the potential to significantly reduce calculation time and make the method widely accessible.  Currently, the main computational challenge is that creating domain sizes typical of optimized devices (\textgreater10~nm) while retaining a high resolution (1~nm) can take a considerable amount of calculation time, especially when a large lattice size is needed.  This challenge has so far greatly limited the ability to systematically study the effects of morphological features on OPV device simulations.  To address these issues, we present a thorough characterization of the morphologies generated using the Ising method, introduce methods to dramatically reduce the calculation time, and show how these changes impact KMC device simulations.

\section{Methods}

\subsection{Ising phase separation algorithm}

When generating a morphology using the Ising model, a three-dimensional lattice is created, and the sites are randomly assigned as either donor or acceptor sites.  Here, a 50:50 blend is implemented, and each site is defined to represent 1~nm$^3$.  Periodic boundary conditions are used in the x and y-directions, and hard boundaries are used in the z-direction to represent a thin film. Next, a simulated phase separation process is executed in which the total energy of the system is allowed to relax over a series of iterations by allowing adjacent sites to be swapped.  

To execute the phase separation process, a pair of adjacent sites with differing types is randomly selected from the lattice.  Then, the total change in energy of the system that would result from swapping them, $\Delta\epsilon$, is calculated and used to determine the probability of the swapping event,
\begin{equation}
P(\Delta\epsilon)=\frac{\exp(-\Delta\epsilon/(kT))}{1+\exp(-\Delta\epsilon/(kT))}.
\end{equation}

Traditionally, the change in energy is calculated by first determining the energy of each site using the Ising Hamiltonian.\cite{watkins2005}  However, here we have developed a mathematically equivalent description that results in a much more computationally efficient algorithm that we have named the bond formation algorithm.  In this algorithm, the swapping process is thought of as the breaking of the bonds present in the initial state and the formation of new bonds in the final state.  In this framework, the change in energy caused by swapping two sites is the difference between the total energy of the initial bonds and the total energy of the final bonds.  To calculate this difference, all that needs to be known is the change in the number of each type of bond between the initial and final states.  Since the method implemented by \citeauthor{watkins2005} only includes interactions between first and second nearest neighbors, the total change in energy is calculated
\begin{equation}
\Delta\epsilon = -\Delta N_1 J-\Delta N_2 \frac{J}{\sqrt{2}},
\end{equation}
where $J$ is the interaction energy, $\Delta N_1$ is the change in the number of first nearest neighbor bonds and $\Delta N_2$ is the change in the number of second nearest neighbor bonds.  A more detailed description of the bond formation algorithm and a comparison to the algorithm based on the traditional energy calculation method is presented in section I of the Supplementary Information.

Once the probability of the swapping event is calculated, a random number generator is used to determine whether the sites are swapped or not.  To continue the phase separation, another suitable pair of sites is randomly chosen and the process is repeated.  Whether the sites are swapped or not, each iteration is counted and the evolution of the system is measured by counting the number of MC steps that have occurred. The number of MC steps is defined as the total number of iterations divided by the total number of sites in the lattice.\cite{watkins2005}  This allows the evolution of the phase separation process to be characterized with a parameter that is independent of the lattice size.

\subsection{Smoothing algorithm}

To modify the morphology, we also implement a smoothing algorithm that removes island sites and smooths rough domain interfaces.  During smoothing, the lattice is scanned one site at a time, and for each site, a roughness factor is calculated.  The roughness factor of a site is calculated by determining the fraction of the 26 total first, second, and third nearest neighbors that are not the same type as the target site.  Island sites and sites at rough domain interfaces are surrounded by mostly sites of the opposite type and will have a large roughness factor.  To smooth the domains, any site that has a roughness factor above a given threshold is switched to the opposite type.  The lattice is continually scanned until all sites are found to have a roughness factor that is below the threshold.  

We have found that a smoothing threshold of 0.52 performs best by reducing the interfacial area without significantly affecting the domain size.  With a 50:50 blend, this smoothing process has an equal probability of smoothing out donor or acceptor sites, and as a result, the blend concentration is not affected.  However, if uneven blend ratios are used, this algorithm does slightly reduce the concentration of the minority component.  A more detailed analysis of the smoothing threshold is presented in section II of the Supplementary Information.

\subsection{Morphology characterization}

Once a morphology was generated, to characterize the average size of the domains, the pair-pair correlation method previously described by \citeauthor{lyons2012} was used.\cite{lyons2012}  The pair-pair correlation function was calculated for each donor site and each acceptor site.  Then, two averages were calculated, one for all donor sites and one for all acceptor sites.  In our pair-pair correlation function algorithm, a resolution of $0.5$~nm was implemented.  As a result, the distance between sites is rounded to the nearest $0.5$~nm.  The average domain size is determined by calculating when the correlation function first crosses over the bulk concentration (0.5).  To calculate the crossover point, a linear interpolation process is used between the two points on either side of the crossover point. 

We note here that in numerous previous studies the domain size has been estimated using the relationship, $d=\frac{3V}{A}$, where $V$ is the volume and $A$ is the interfacial area.\cite{marsh2007,groves2008a,groves2008b,groves2009,groves2010a,meng2010,feron2012a,maqsood2013}  This relationship is only strictly valid when the domains are spherical, and since the Ising model produces highly non-spherical domains, this approximation severely overestimates the domain size.  We estimate that these studies have likely overestimated the domain size present in their morphologies by about 75\%.  A more detailed comparison between these methods in presented in section III of the Supplementary Information.

To characterize the shape and connectivity of the domains, the interfacial area to volume ratio and tortuosity was calculated. The interfacial area to volume ratio was calculated by counting the number of cubic site faces between a donor and an acceptor site and then dividing the total count by the total number of sites in the lattice.  The tortuosity is defined for an individual site as the length of the shortest available path from the given site through the same domain type to the collecting electrode divided by the length of the corresponding shortest straight path.\cite{wodo2012}  To calculate this, a three-dimensional, breadth-first search, graph traversal method was used to determine the shortest path from all donor and acceptor sites to their respective collecting electrode.  The tortuosity was then calculated for all donor sites at the cathode interface and all acceptor sites at the anode interface to give a data set that is representative of the charge transport paths through the entire thickness of the film.  Since a 50:50 blend is studied here, the donor and acceptor paths should be statistically equal and were averaged together.

To determine how the morphological changes impact simulated device performance, two simple KMC simulation benchmarks were used. A detailed description of our KMC simulation methods can be found in our previous work.\cite{heiber2012}  In all KMC simulations, an uncorrelated Gaussian density of states was implemented with a standard deviation of 75~meV, and a temperature of 300~K was used.  First, an exciton quenching efficiency test was performed by generating excitons with a diffusion length of 10~nm and a lifetime of 500~ps.\cite{mikhnenko2013}  In this test, all excitons that reach the donor-acceptor interface within their lifetime are dissociated, and those that do not reach the interface relax to the ground state. The interaction distance for exciton dissociation was set to 2~nm in accordance with our previous studies.\cite{heiber2013c} Exciton quenching tests were performed for 1000 excitons on 9 different energetic disorder configurations for each morphology. Second, the effect of the morphology on charge transport was probed using a simulated thin film time of flight (ToF) experiment.  In this test, a hole is created at a randomly selected donor site at one surface of the lattice and allowed to undergo standard hopping behavior under an applied electric field.  When it reaches the opposite surface, it is removed from the lattice, the transit time is recorded, and the entire process is repeated.  With an applied field of $10^{7}$~V/m, ToF simulations were performed for 1000 carriers on 9 different energetic disorder configurations for each morphology.

Additional computational details and calculation time benchmarks for the morphology sets are provided in section IV of the Supplementary Information.  An open-source software package for supercomputer use and a more simple web-based morphology generation tool to create morphologies using the methods described here are available online.\cite{heiber2014a} 

\section{Results and Discussion}

\subsection{Effect of the interaction energy}

The first and most important behavior to understand is the effect of the interaction energy, $J$, on the generated morphologies.  Previous studies have often used an interaction energy of $1~kT$,\cite{watkins2005,meng2010,meng2011,feron2012a,feron2012b} or have neglected to specify the interaction energy used, without discussing the effect of changing the interaction energy.  Our preliminary tests indicated that domain growth was much faster when using smaller interaction energies.  This finding prompted an investigation as to whether or not a smaller interaction energy could be used to more efficiently generate model morphologies with large domain sizes.  To characterize the effect of the interaction energy in greater detail, 24 independent morphologies were generated on a 50 by 50 by 50 lattice for $J=0.4$, $0.6$, $0.8$, and $1.0~kT$, varying the number of iterations (MC steps) to create domain sizes in the range of 5 to 10~nm.  The domain size, interfacial area to volume ratio, and tortuosity were calculated for each morphology.  In the following sections, the data points in the figures indicate the mean of each data set, and the error bars represent one standard deviation.  

\begin{figure}[h]
\includegraphics[]{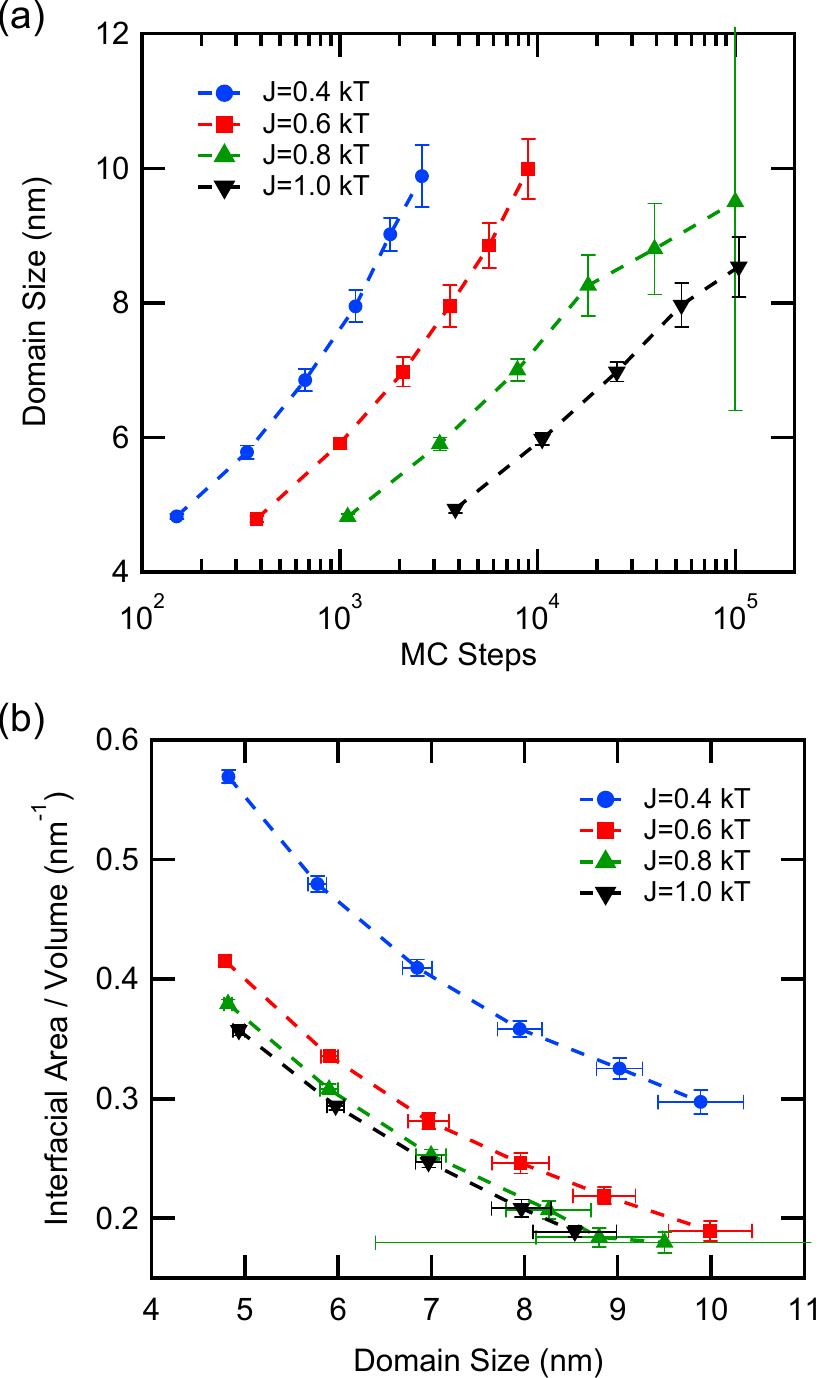}
\caption{\label{fig:fig1} The effect of the interaction energy, $J$, on morphology generation. (a) Growth of domains with increasing MC steps and (b) resulting interfacial area to volume ratio for $J=0.4~kT$ (blue circles), $J=0.6~kT$ (red squares), $J=0.8~kT$ (green triangles), and $J=1.0~kT$ (black inverted triangles).}
\end{figure}

Figure \ref{fig:fig1}a shows how the domain size grows during the simulated phase separation process as a function of the number of MC steps for different magnitudes of interaction energy.  In all cases, the domain growth is fast initially and then slows down over time.  This slowing of domain growth over time is particularly pronounced at higher interaction energies, as described previously by \citeauthor{binder1974b}.\cite{binder1974b} However, the number of MC steps required to reach a specific domain size varies dramatically, with about one and a half orders of magnitude difference between interaction energies of $0.4$ and $1.0~kT$.  Most significantly, this leads to much longer computational time.  This behavior can be explained by considering the effective mobility of the sites in the lattice.  When the interaction energy is large, it is much less likely for sites to go through the energetically unfavorable intermediate states that are required for site rearrangement and eventual domain growth.  

In addition, the variability of the domain size obtained for specific number of MC steps increases as the domains grow in size.  This general trend is present for all interaction energies tested but appears to be reduced slightly when using a smaller interaction energy.  However, with an interaction energy of $0.8~kT$, the domain sizes appear highly varied once an average domain size of 9~nm is reached.  This suggests that the lower interaction energies produce morphologies with domains that are more uniform in size.

\begin{figure}[h]
\includegraphics[]{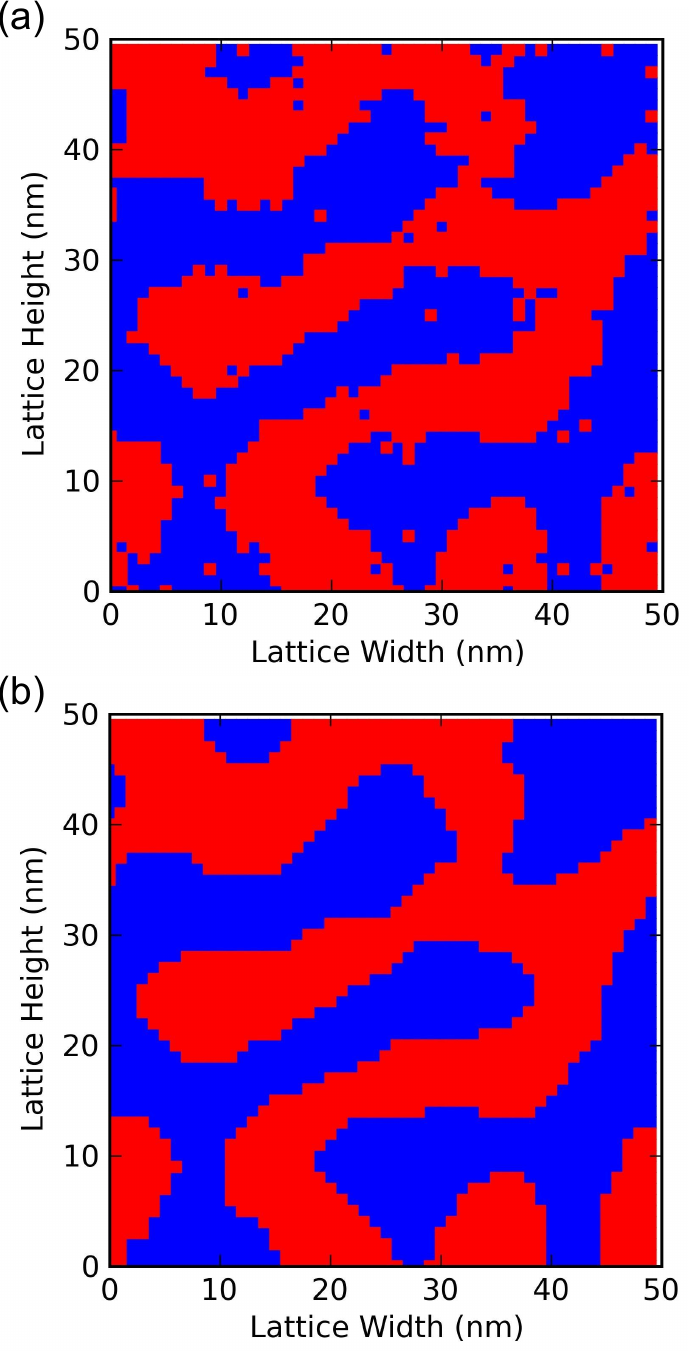}
\caption{\label{fig:fig2} The effect of smoothing on the generated morphologies. Cross-sectional images for $J=0.4~kT$ (7.0~nm domains after 750 MC steps) (a) without smoothing and (b) after smoothing.}
\end{figure}

To compare the generated morphologies in more detail, the interfacial area to volume ratio was calculated and is shown in Fig.~\ref{fig:fig1}b.  If the domains are  shaped differently, the interfacial area to volume ratio should be affected.  For example, if the domains tend to be more spherical in shape, the interfacial area to volume ratio will be lower than if the domains tend to be more cylindrical.  It is very clear that the morphologies generated with an interaction energy of $0.4~kT$ have a much larger interfacial area to volume ratio.  To visualize this difference, Figure \ref{fig:fig2}a shows a cross-sectional image of a morphology generated with an interaction energy of $0.4~kT$.  It is clear that this morphology has quite a few island sites and very rough interfaces that contribute to the large interfacial area to volume ratio.  However, it is still unclear if the underlying domain shape is significantly different.

In an attempt to investigate the potential domain shape differences apart from the effects of island sites and rough domain interfaces, the smoothing algorithm described in the Methodology section is applied.  Figure \ref{fig:fig2} shows how the smoothing algorithm modifies the morphology when using an interaction energy of $0.4~kT$.  From these cross-sectional images, it is clear that the smoothing algorithm successfully removes all island sites and smooths rough domain interfaces without significantly changing the size or shape of the domains.

\begin{figure}[h]
\includegraphics{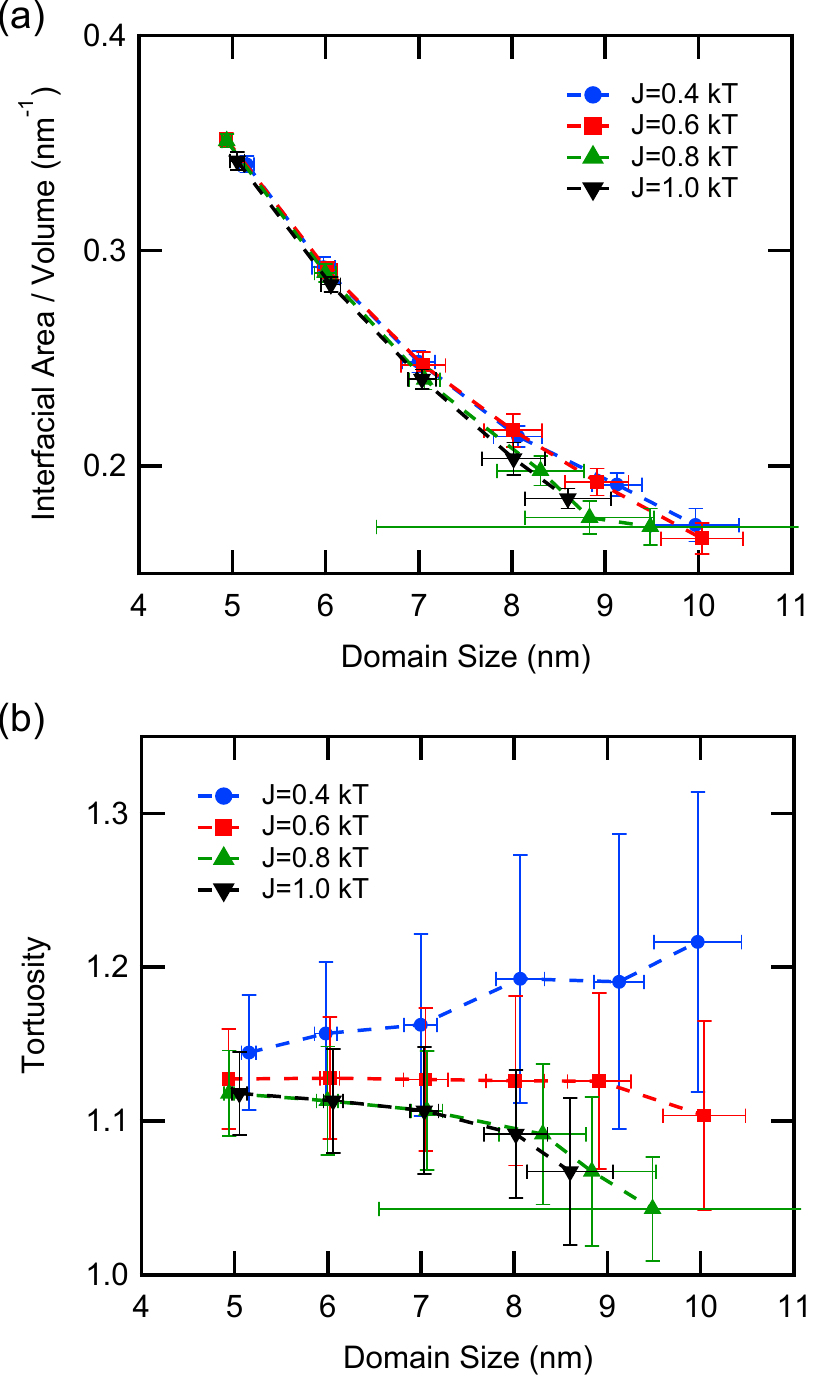}
\caption{\label{fig:fig3}The effect of the interaction energy, $J$, on final smoothed morphologies. (a) Interfacial area to volume ratio and (b) tortuosity as a function of domain size after smoothing for $J=0.4~kT$ (blue circles), $J=0.6~kT$ (red squares), $J=0.8~kT$ (green triangles), and $J=1.0~kT$ (black inverted triangles).}
\end{figure}

To analyze and characterize the effect of smoothing, all previously generated morphology sets were smoothed and the domain size, interfacial area to volume ratio, and tortuosity were recalculated.  The resulting data, shown in Figure \ref{fig:fig3}a, indicates that once smoothing is applied, the high interfacial area to volume ratio of morphologies generated with low interaction energies is greatly reduced, approaching the values originally attained with higher interaction energies.  The smoothing process has very little impact on the morphologies generated with $J=1.0~kT$.  After smoothing, there is much less difference in the interfacial area to volume ratio between the morphologies generated with different interaction energies at any given domain size.  However, significant differences do start to arise when the domains reach about 8~nm or larger.  In this regime, interaction energies of 0.8 and $1.0~kT$ produce domains with a significantly smaller interfacial area to volume ratio than both $0.6~kT$ and $0.4~kT$.

Figure \ref{fig:fig3}b shows the tortuosity for the smoothed morphologies.  Similar to the trend observed with the interfacial area to volume ratio, the tortuosity obtained with each interaction energy is initially very similar but deviates as the domains grow in size.  In particular, different trends are observed for each interaction energy.  For 0.8 and $1.0~kT$, the tortuosity decreases as the domains grow in size.  For $0.6~kT$, the tortuosity remains almost constant as the domains grow in size.  And for $0.4~kT$, the tortuosity increases as the domains grow in size.  As a result, each interaction energy produces morphologies that have distinct differences and would be expected to produce different KMC simulation results.  These potential effects on KMC simulations are tested and discussed in the section D.

From these tests, it appears that using an interaction energy of $0.6~kT$ is best for generating controlled model BHJ morphologies.  The domains grow much faster than with higher interaction energies, which reduces computational time, and the domains are also more uniform in size.  In addition, the tortuosity of the morphology is fairly constant as the domains grow in size, which allows one to look at the impact of the domain size independent from the tortuosity.  We will also show in the next subsection that this constant tortuosity works especially well with the lattice rescaling method used to efficiently create morphologies with larger domain sizes.  As a result, the subsequent section focuses only on morphologies generated using an interaction energy of $0.6~kT$.

\subsection{Simplifications for computational efficiency}

Regardless of the magnitude of the interaction energy, there are several methods for significantly reducing the computation time required to generate a particular model morphology.  The first method is to reduce the lateral dimensions of the lattice.  Because periodic boundary conditions are used in the plane of the film, the choice of lateral dimensions is somewhat arbitrary.  However, the number of sites in the lattice will affect the calculation time per MC step.  As a result, it is common to use lateral dimension that are smaller than the thickness dimension.  Reducing two of the dimensions of the lattice can significantly reduce the total number of sites.  For all previous tests a 50 by 50 by 50 site lattice was used, but equivalent morphologies can be generated using a smaller lattice.  However, at some point it is also expected that too small of a lattice may introduce confinement effects that change the domain size and/or domain shape.

\begin{figure}[h]
\includegraphics[]{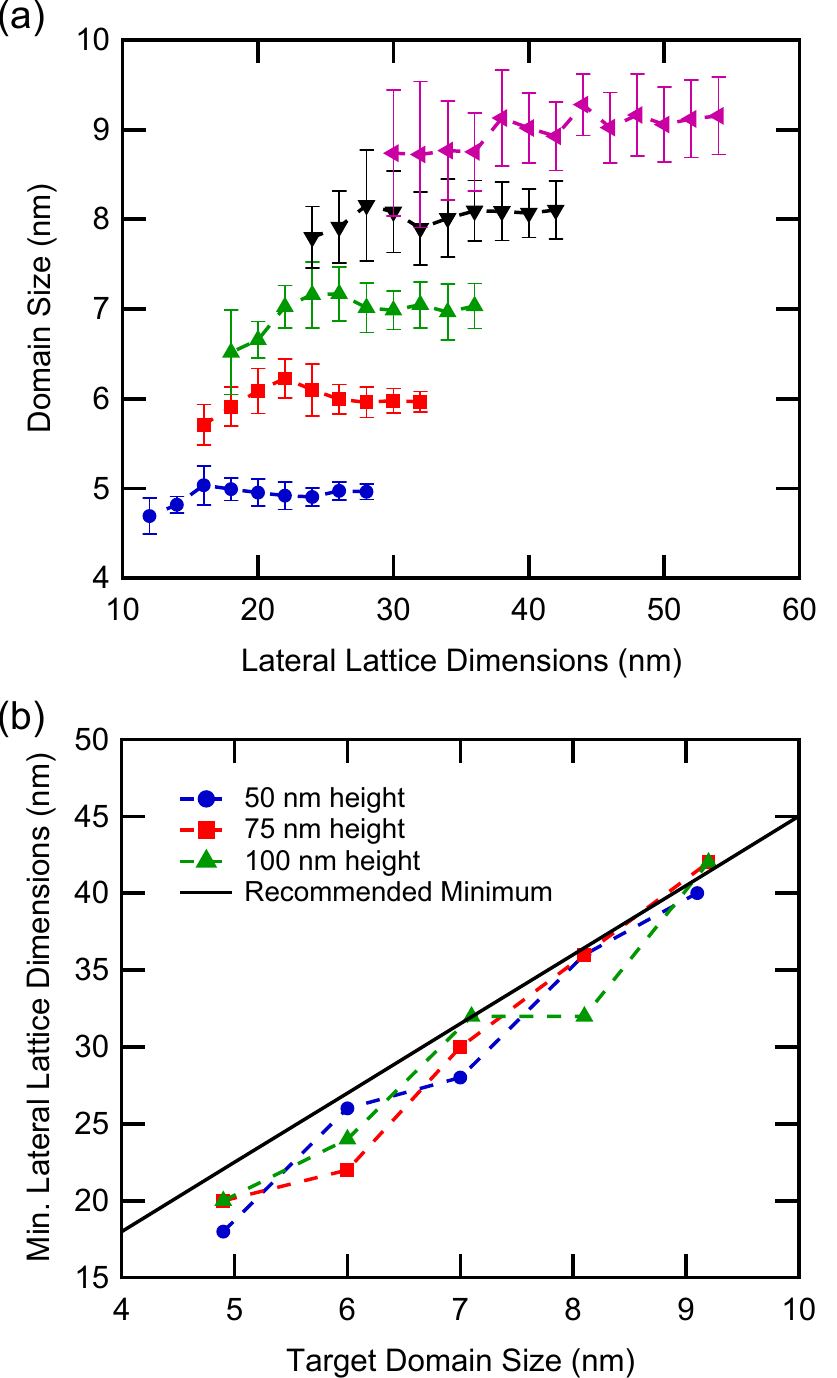}
\caption{\label{fig:fig4} The effect of the lateral lattice dimensions on morphologies generated with $J=0.6~kT$. (a) The effect of the lateral lattice dimensions on target domain sizes of 5, 6, 7, 8, and 9~nm using a lattice height of 50~nm and (b) the minimum lateral lattice dimensions as a function of target domain size for lattice heights of 50, 75, and 100~nm.  The solid black line shows the recommended minimum lateral dimensions for the desired target domain size corresponding to 4.5 times the target domain size.}
\end{figure}

To probe this behavior, morphologies were created with domain sizes ranging from 5 to 9~nm on lattices where the length and width (lateral dimensions) were varied but always equal using lattice heights of 50, 75, and 100~nm.  20 independent morphologies were created for each combination of domain size and lattice size, and after executing the swapping and smoothing algorithms, the final domain size was calculated and recorded for each morphology.  Figure \ref{fig:fig4}a shows how the domain size is affected by decreasing the lateral lattice dimensions when using a lattice height of 50~nm.  When the lateral lattice dimensions are very large, the domain size obtained does not depend on the lattice dimensions as expected.  However, as the lateral dimensions become smaller, changes to the domain size eventually start to be observed due to lattice confinement effects.  The onset of confinement effects occurs at larger lateral dimensions when creating larger domains, and this same trend was observed for lattice heights of 75 and 100~nm (not pictured).

To characterize this relationship in more detail, for each target domain size and lattice height tested, the lateral dimensions at which the domain size started to be noticeably affected was recorded.  We define this transition point as the minimum lateral dimensions.  Figure \ref{fig:fig4}b shows how the minimum dimensions change as a function of the target domain size for each lattice height tested.  We found that the onset of lattice confinement effects is not dependent on the lattice height, and that as a general rule, as long as the lateral dimensions are greater than or equal to 4.5 times the target domain size, lattice confinement does not significantly impact the final morphology.  As a result, the calculation time can be reduced by using smaller lateral dimensions, but the lattice size can only be safely reduced down to 4.5 times the desired domain size.  We have observed a similar limit when using other interaction energies, but have not performed a detailed characterization of these additional cases.  

The methods described so far work well for creating relatively small domains, but as the domains continue to grow in size, the rate of domain growth also decreases, as discussed previously. The final way to reduce the calculation time is to utilize a lattice rescaling method, as used by \citeauthor{mcneill2007}\cite{mcneill2007}  This method essentially stretches the lattice equally in all three dimensions, making both the lattice and the domains larger without altering the shape and connectivity of the domains. For creating domains that are larger than 10~nm, which is more typical of many BHJ blend materials, a lattice rescaling method can dramatically reduce the calculation time.  As an example, to create a morphology that is representative of a 100~nm film with 16~nm domains, an 80 by 80 by 100 lattice is needed.  Even when using a lower interaction energy, without the rescaling method, this calculation takes several days to create a single morphology on one processor.  With the rescaling method, a morphology with 8~nm domains can be created on a much smaller 40 by 40 by 50 lattice and then rescaled by a factor of 2 to obtain the final desired morphology in about 2 hours.  Additionally, a rescaling factor of 3 could be used on the same initial morphology to create 24~nm domains with only slightly more calculation time.

\begin{figure}[h]
\includegraphics{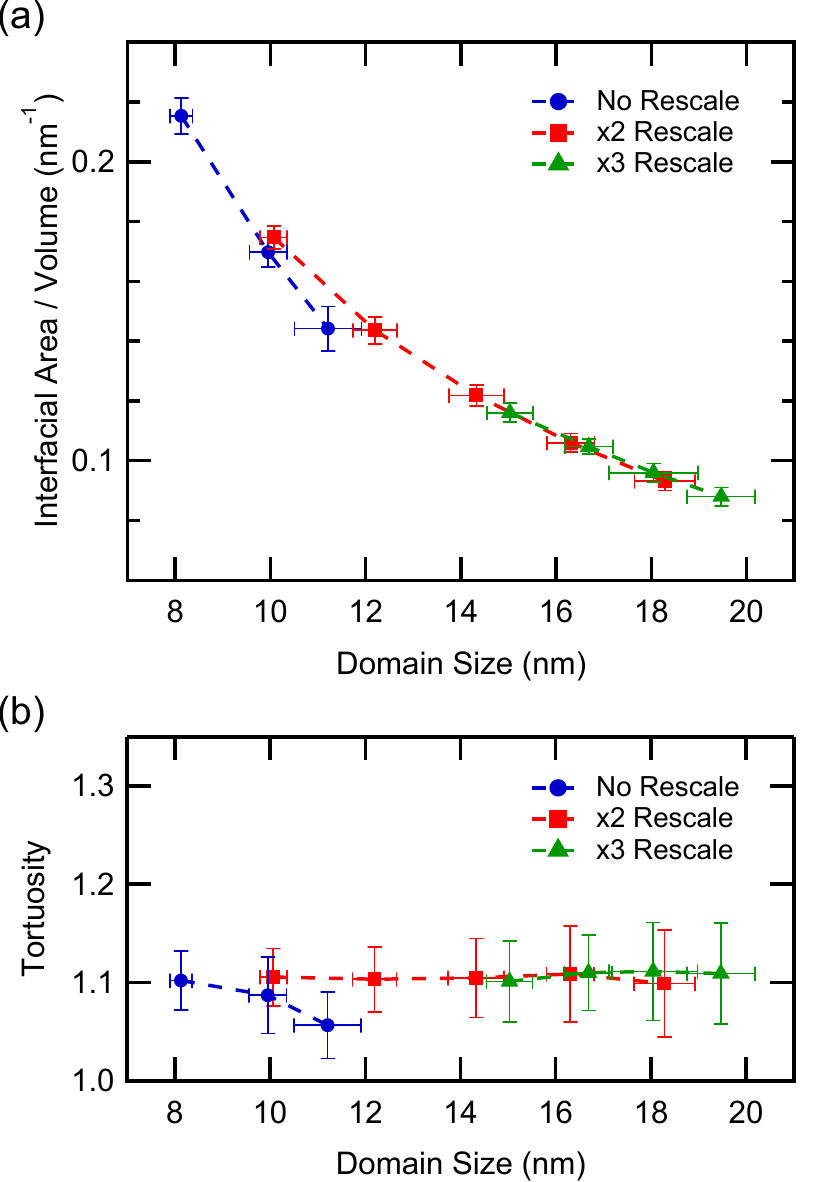}
\caption{\label{fig:fig5} Effect of lattice rescaling on morphologies.  (a) Interfacial area to volume ratio and (b) tortuosity as a function of domain size for $J=0.6~kT$ without rescaling (blue circles), x2 rescaling (red squares), and x3 rescaling (green triangles).}
\end{figure}

However, it is also a concern if the rescaling method introduces major changes to the tortuosity.  To characterize this, 24 independent morphologies were created with an interaction energy of $0.6~kT$ on a lattice with a 100~nm height for a range of domain sizes using both the normal method and the rescaling method.  Lateral lattice dimensions were set to 4.5 times the target domain size, and smoothing was applied before and after rescaling.  Figure \ref{fig:fig5} shows that without rescaling, both the interfacial area to volume ratio and the tortuosity begin to decrease for domain sizes over 10~nm.  However, with the rescaling method, when initial domain sizes are created in the range of 5 to 9~nm, this does not occur and both the interfacial area to volume ratio and the tortuosity remain constant.  If the rescaling method is used with interaction energies or domain size ranges where the tortuosity is not constant, the tortuosity will fluctuate when switching between the normal method and each rescaling factor.  However, with an interaction energy of $0.6~kT$ and initial domain sizes in the range of 5 to 9~nm, the rescaling method can be safely used to efficiently create morphologies with a wide range of domain sizes.

\subsection{Impacts on KMC device simulations}

In section A, we showed how the interaction energy can impact the simulated phase separation process, yielding morphologies with different domain shapes characterized by differences in the interfacial area to volume ratio and tortuosity.  To determine how these differences would impact device simulations, we have generated morphology sets similar to typical OPV devices and performed two benchmark KMC simulations. Interaction energies of 0.4, 0.6, and 0.8$~kT$ were used to create morphology sets with domain sizes of approximately 15, 18, and 21~nm on lattices representing a film thickness of 102~nm.  A total of 9 morphology sets with 24 morphologies each were generated.  Domain smoothing, minimum lateral dimensions, and lattice rescaling were used, as described previously, to quickly generate the final morphologies.  Instead of taking several days to generate each morphology, each morphology was created on one processor in only 1-2 hours.  Additional characterization of these morphology sets is shown in section V of the Supplementary Information.

\begin{figure}[h]
\includegraphics{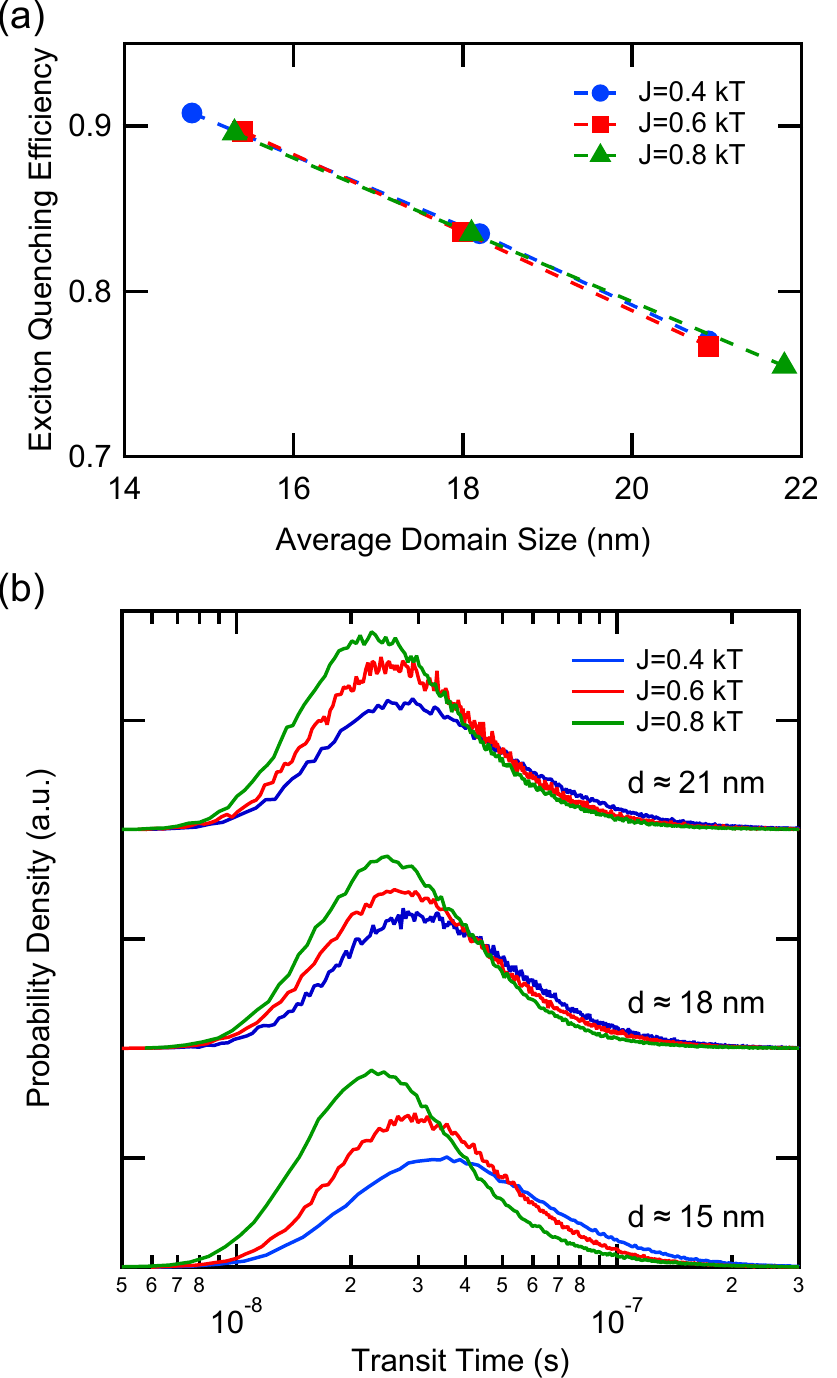}
\caption{\label{fig:fig6}The effect of the interaction energy, $J$, on KMC device simulation performance using model BHJ morphologies. (a) Exciton quenching efficiency and (b) transit time distributions for $J=0.4~kT$ (blue circles), $J=0.6~kT$ (red squares), and $J=0.8~kT$ (green triangles) and domain sizes of 15, 18, and 21~nm.}
\end{figure}

First, the simulated exciton quenching efficiency is shown in Fig.~\ref{fig:fig6}a, and we find that while the exciton quenching efficiency is dependent on the domain size as expected, the interaction energy used to generate the morphology has almost no impact.  As a result, for studies focused on modeling exciton diffusion and dissociation at the donor-acceptor interface in BHJ devices, the choice of interaction energy is not very significant as long as domain smoothing is applied to remove the island sites that will act as exciton quenching sites.  

However, the situation is significantly different for charge transport simulations.  The transit time distributions resulting from time of flight charge transport simulations are shown in Fig.~\ref{fig:fig6}b.  It is clear that for all three domain sizes tested, a smaller interaction energy results in longer transit times and a more dispersive distribution.  Looking back at Fig.~\ref{fig:fig3}b, we observed noticeable differences in tortuosity for each of these interaction energies.  In general, the tortuosity increased when decreasing the interaction energy. This same trend persists for the larger domain sizes generated here and is shown in section V of the Supplementary Information.  This increase in tortuosity appears to significantly slow down the charge transport and increase the dispersion.  We expect this effect to be enhanced when simulating thicker films or when the electric field is weaker.  As a result, for simulation and modeling studies in which charge transport is an important factor, the choice of interaction energy will have a significant impact.  For example, in a full device simulation, slower transport should lead to reduced charge collection efficiency and increased charge recombination.

\section{Conclusions}

Overall, we have provided a detailed characterization of the Ising method for generating model bulk heterojunction (BHJ) morphologies.  We have investigated the effect of the interaction energy and demonstrated several methods for reducing the computation time required to generate model morphologies.  We first introduced a new algorithm called the bond formation algorithm for calculating the site swapping probabilities, which gave rise to a major increase in calculation speed.  We then demonstrated how a smaller interaction energy of $0.6~kT$, when used with a smoothing algorithm, produces pure domains with faster domain growth than previous work and a tortuosity that is almost independent of the domain size.  Next, we probed the limits of using small lateral lattice dimensions with periodic boundary conditions to reduce the calculation time. Finally, we characterized the performance of a lattice rescaling method to be used when creating large domains (\textgreater 10~nm) and identified the conditions that allow for the creation of a wide range of domain sizes.  In total, these developments reduce the morphology generation time by several orders of magnitude.

Combining all methods discussed here, morphologies with domain sizes and thicknesses typical of optimized BHJ OPVs were able to be efficiently generated for kinetic Monte Carlo (KMC) simulations.  We have shown how changes in domain size and tortuosity can significantly impact charge transport, which can have a broad impact on charge recombination and ultimately the power conversion efficiency.  With this in mind, it is imperative that future modeling studies are precise and forthcoming regarding the methods used for morphological modeling.  In particular, studies should pay close attention to how the domain size is determined, how the tortuosity changes with increasing domain size, and how the lattice size is chosen.

Including detailed morphological features into device models continues to be an important step towards the ability to accurately analyze, simulate, and ultimately predict device performance. 
The advancements described here have been implemented and published in an open-source software code for supercomputer use\cite{heiber2014a} and in a user-friendly web-based software tool.\cite{heiber2014b}  With the methods and morphology generation tools freely available, other researchers can now easily generate model (BHJ) morphologies in a computationally efficient manner and apply them to novel systematic device modeling efforts.  These developments will allow KMC simulations to be readily performed on large sets of morphologies created with a wide range of parameters, leading to increased understanding of the link between morphology and device performance.

\begin{acknowledgments}
We thank the LORD corporation and the National Science Foundation grant NSF-DMR 0512156 for funding and Prof. Mesfin Tsige and Gary Leuty for discussions and simulation help. We also thank Prof. Carsten Deibel for his support and insight.  Computational resources were provided by Prof. Mesfin Tsige at The University of Akron and by the Department of Physics and Astronomy at the University of W\"urzburg.
\end{acknowledgments}

\bibliography{references}

\end{document}


\title{Supplementary Information for Efficient Generation of Model Bulk Heterojunction Morphologies for Organic Photovoltaic Device Modeling}

\author{Michael C. Heiber}
\email{michael.heiber@physik.uni-wuerzburg.de}
\affiliation{Department of Polymer Science, The University of Akron, Goodyear Polymer Center, Akron, Ohio 44325, USA}
\affiliation{Experimental Physics VI, Julius-Maximilians-University of W{\"u}rzburg, Am Hubland, D-97074 W{\"u}rzburg, Germany}

\author{Ali Dhinojwala}
\email{ali4@uakron.edu}
\affiliation{Department of Polymer Science, The University of Akron, Goodyear Polymer Center, Akron, Ohio 44325, USA}

\date{\today}

\pacs{}

\maketitle

\section{Ising phase separation algorithm details and analysis}

As discussed in the main text, in the Ising method, a simulated phase separation process is executed in which the total energy of the system is allowed to relax over a series of iterations by swapping adjacent sites.  Computationally, there are several ways to implement the site swapping process and, in particular, the energy change calculation, which is the most computationally intensive step.  Previous studies have not explained the computational algorithms used in detail, so we have developed and optimized two possible algorithms, which are characterized here.  

The first method, which we have named the site energy algorithm, is a fairly straightforward method that emerges naturally from using the Ising Hamiltonian to determine the energy of the system.  In this method, the energy of each site is calculated,
\begin{equation}\label{eqn:ising}
\epsilon_i = -\frac{J}{2(d_{ij}/a)}\sum_j(\delta_{t_i,t_j}-1),
\end{equation}
where $J$ is defined as the interaction energy parameter, which is the difference between the donor-donor or acceptor-acceptor interaction energy and the donor-acceptor interaction energy, $d_{ij}$ is the distance between site $i$ and site $j$, $a$ is the lattice constant, $\delta_{t_i,t_j}$ is the Kronecker delta, and $t_i$ and $t_j$ are the types of sites $i$ and $j$, respectively.\cite{watkins2005} The summation over $j$ includes only first nearest neighbors at a distance of $a$ from site $i$ and second nearest neighbors at a distance of $\sqrt{2}a$.  By convention, interactions between third-nearest neighbors and greater are not included.  It is assumed that the donor-donor interaction energy and the acceptor-acceptor interaction energy are equal, which simplifies the system to a single interaction energy parameter.  In addition, the interaction energy, $J$, is typically defined in units of $kT$, and a positive interaction energy indicates that donor-donor interactions and acceptor-acceptor interactions are stronger than donor-acceptor interactions.  This favorable like-like interaction drives the phase separation process.

In the site energy algorithm, first, the energy of each site in the lattice is calculated using Eqn.~\ref{eqn:ising}.  Then, when the swapping pair is chosen, the energies of all first, second, and third nearest neighbors of each pair site are added together to calculate the initial local energy.  Next, the sites are temporarily swapped, and the energy of all nearby sites affected by the swap are recalculated using Eqn.~\ref{eqn:ising} and stored in a temporary data structure.  Finally, the newly calculated energies of all first, second, and third nearest neighbors of each pair site are again added together to calculate the final local energy.  The difference between the initial and final local energies is the total change in energy for the swapping event, and the probability calculation then determines whether to accept or reject the swapping event.  If accepted, the final site energies stored in the temporary data structure are applied to the lattice.  If the swap is rejected, the temporarily swapped sites are switched back to their original type.  

Even in a very highly optimized implementation of this algorithm, we have observed that the site energy algorithm described here is very slow.  After careful consideration and profiling of the site energy algorithm, the main problem with this algorithm is that for each iteration, a large number of sites must have their energies recalculated, and the site energy calculation step itself is fairly slow due to the summation over $j$, which includes 18 neighboring sites.  This algorithm is very computationally wasteful because, when calculating the final energy of each local site after the temporary swap, at most only 2 of the 18 sites have changed. As a result, much of the loop over $j$ sites is unnecessary but is performed numerous times for each iteration. 

To significantly reduce the wasted computations and improve the calculation speed, we have developed a mathematically equivalent concept and algorithm referred to here as the bond formation algorithm. Alternatively, the swapping process is thought of as the breaking of the bonds present in the initial state and the formation of new bonds in the final state.  In this framework, the change in energy caused by swapping is the difference between the total energy of the initial bonds and the total energy of the final bonds.  To calculate this difference, all that needs to be known is the change in the number of each type of bond between the initial and final states.  

The Ising Hamiltonian only counts like-like interactions between first and second-nearest neighbors, so in the bond formation algorithm, only like-like bonds are considered.  For example, if site $i$ has $m$ first nearest neighbors that are of the same type, then $m$ first-nearest neighbor bonds will be broken when the sites are swapped.  However, given that there are 5 first-nearest neighbors after excluding site $j$, swapping will cause site $i$ to form $5-m$ new bonds in the final state, resulting in a net change of $5-2m$ first nearest neighbor bonds.  If site $j$ has $n$ first nearest neighbors that are of the same type, similarly, the net change for site $j$ is $5-2n$ first nearest neighbor bonds, resulting in a total change of $10-2m-2n$ first nearest neighbor bonds.  An analogous calculation is also performed for the breaking and formation of the potential 12 second nearest neighbor bonds.  As a result of this simplification, the total change in energy is calculated
\begin{equation}
\Delta\epsilon = -\Delta N_1 J-\Delta N_2 \frac{J}{\sqrt{2}},
\end{equation}
where $\Delta N_1$ is the change in the number of first nearest neighbor bonds and $\Delta N_2$ is the change in the number of second nearest neighbor bonds.  This calculation is equivalent to the site energy algorithm but can be executed much faster.  In addition, the energy of all sites does not need to be calculated at any point.

\begin{figure}[h]
\includegraphics[]{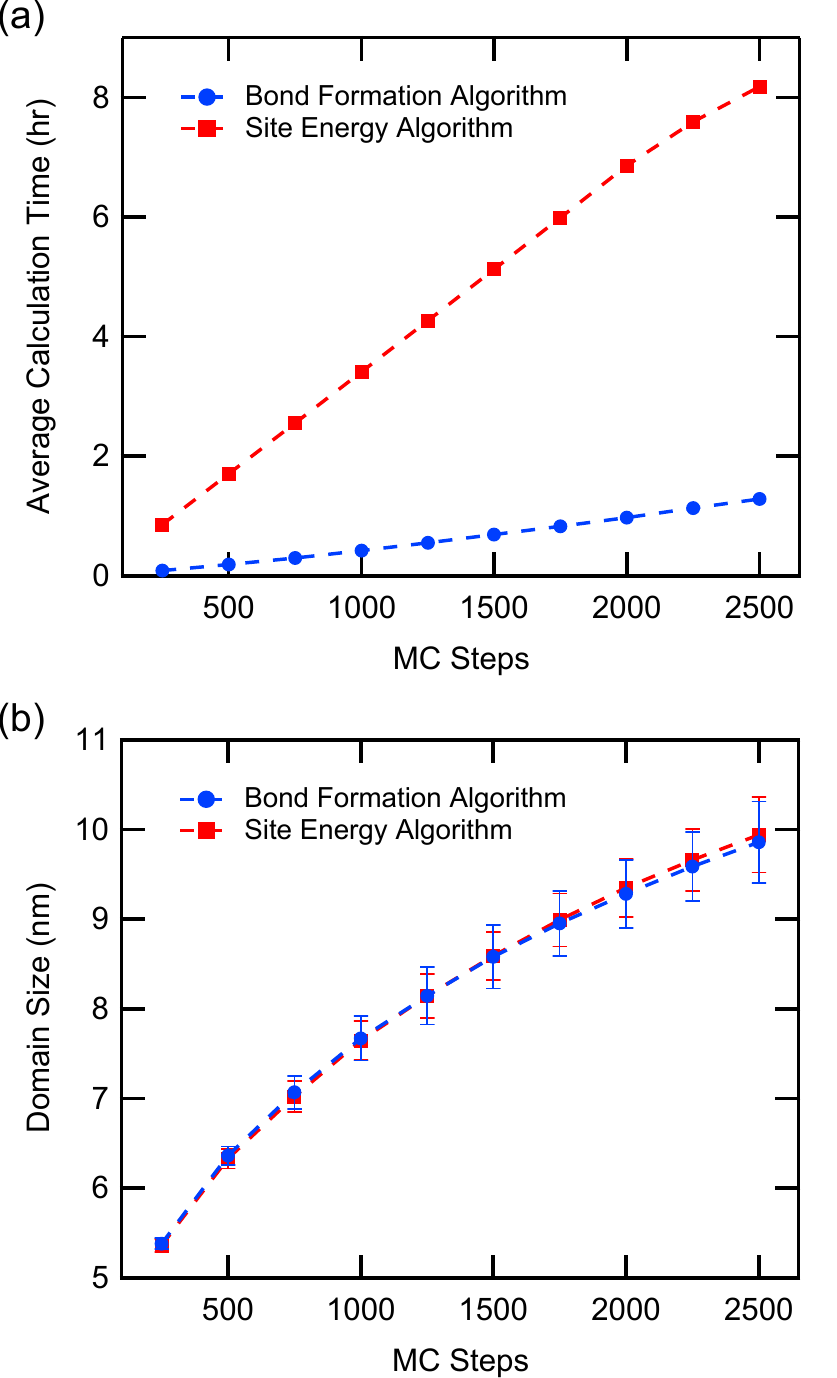}
\caption{\label{fig:figS1} Algorithm calculation time analysis. (a) MC steps vs. calculation time and (b) MC steps vs. domain size for the bond formation algorithm (blue circles) and the site energy algorithm (red squares) with $J=0.4~kT$.}
\end{figure}

To compare the two algorithms, each algorithm was independently run 24 times for 2500 MC steps on a 50 by 50 by 50 lattice with an interaction energy of $0.4~kT$.  After every 250 MC steps, the domain size, interfacial area to volume ratio, and elapsed calculation time were determined.  The average and standard deviation of each parameter were calculated at each MC step interval for the 24 runs.  The calculation speed difference between the site energy algorithm and the bond formation algorithm is shown in Figure \ref{fig:figS1}a.  The bond formation algorithm was about 7-8 times faster than the site energy algorithm.  In both cases, the calculation time is approximately linearly proportional to the number of MC steps executed, indicating that the swapping portion of the calculation takes the majority of the time instead of the domain size calculation.  To demonstrate that both algorithms produce the same phase separation behavior, Figure \ref{fig:figS1}b shows how the domain size grows as a function of the number of MC steps. Both algorithms produce identical results within the expected variability due to the random nature of the morphology generation process.

\section{Smoothing Algorithm Analysis}

In the main text, we have described a smoothing algorithm that removes island sites and smooths rough domain interfaces.  During this process, the lattice is scanned one site at a time, and for each site, a roughness factor is calculated.  The roughness factor of a site is calculated by determining the fraction of the 26 total first, second, and third nearest neighbors that are not the same type as the target site.  Island sites and sites at rough domain interfaces are surrounded by mostly sites of the opposite type and will have a large roughness factor.  To smooth the domains, any site that has a roughness factor above a given threshold is switched to the opposite type.  The lattice is continually scanned until all sites are found to have a roughness factor that is below the threshold.  

\begin{figure}[h]
\includegraphics[]{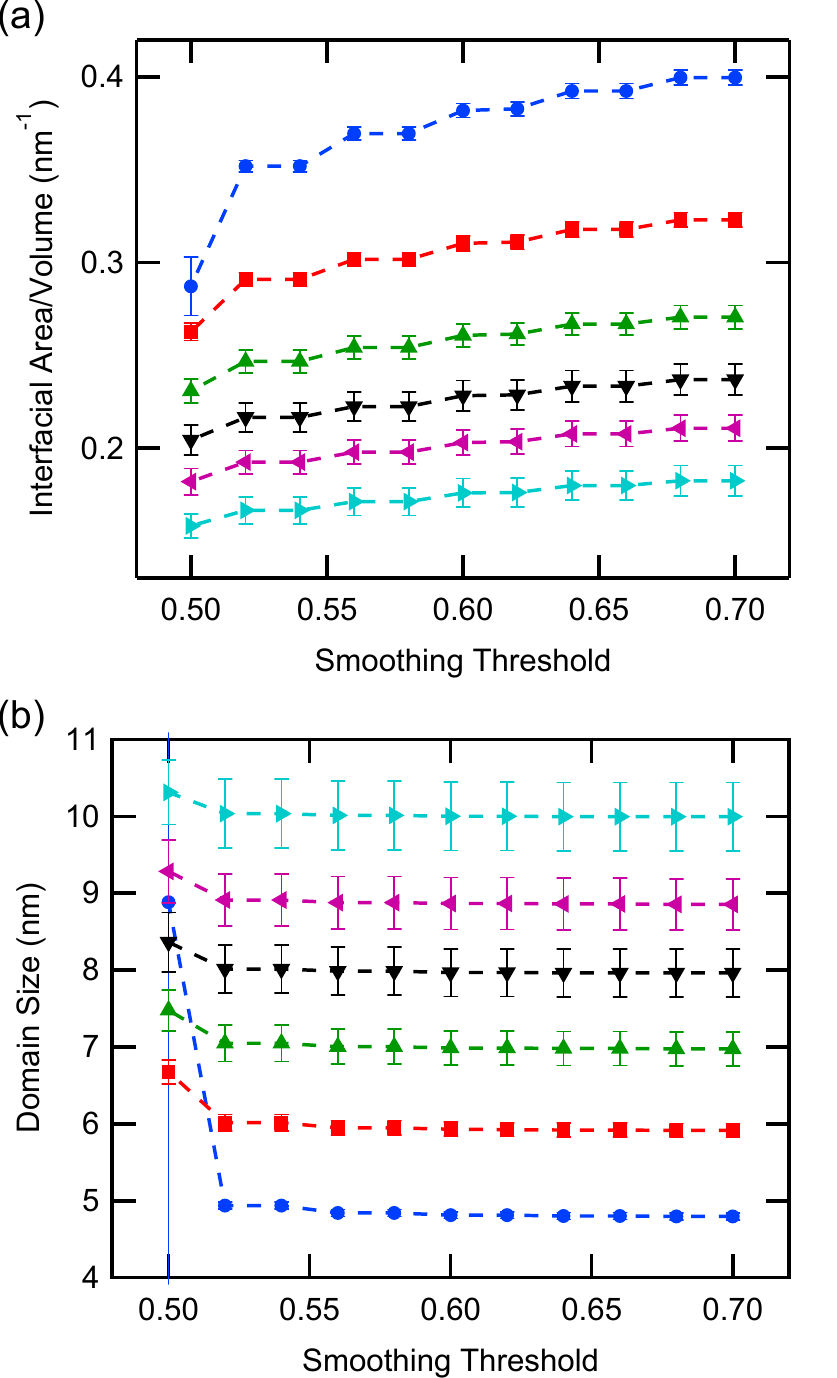}
\caption{\label{fig:figS2} Smoothing algorithm analysis. The effect of the smoothing threshold on the (a) interfacial area to volume ratio and (b) domain size for morphologies with target domain sizes of 5, 6, 7, 8, 9, and 10~nm created with $J=0.6~kT$.}
\end{figure}

To test the effect of the smoothing threshold on the resulting morphology, the smoothing process was applied with a range of values for the smoothing threshold on the morphology sets previously generated with domain sizes of 5, 6, 7, 8, 9, and 10~nm on a 50 by 50 by 50 lattice with $J=0.6~kT$.  After each smoothing treatment, the domain size and interfacial area to volume ratio of each final morphology were calculated and recorded.  Each morphology set corresponding to a specific smoothing threshold and domain size consisted of 24 morphologies, and from these, the average and standard deviation of each characterizing parameter were determined.  Figure \ref{fig:figS2}a shows how the interfacial area to volume ratio decreases as the smoothing threshold decreases for all domain sizes.  As the smoothing threshold decreases, the algorithm becomes more and more aggressive at removing island sites and sites at rough interfaces.  

The effect of the smoothing process on the domain size is shown in Fig.~\ref{fig:figS2}b.  Here, no change in the domain size over the majority of the range of smoothing thresholds tested is observed, even though the interfacial area changes significantly over the same range.  This ensures that the smoothing process is in fact only removing island sites and rough interfacial sites without impacting the underlying domains.  However, it can be seen that at a smoothing threshold below 0.52, the domain size starts to become affected.  As a result, this transition point represents the limit where the smoothing algorithm becomes too aggressive and begins modifying the underlying domains.  This limit appears to be equal for all target domain sizes tested.  The only minor deviation is that for a starting domain size of 5~nm, the domain size is slightly modified at larger smoothing thresholds.  This deviation increases even further for domain sizes smaller than 5~nm.  As a result, this smoothing algorithm can only be safely applied on morphologies with domain sizes greater than or equal to 5~nm.  

\section{Domain Size Calculation Comparison}

\begin{figure}[h]
\includegraphics[]{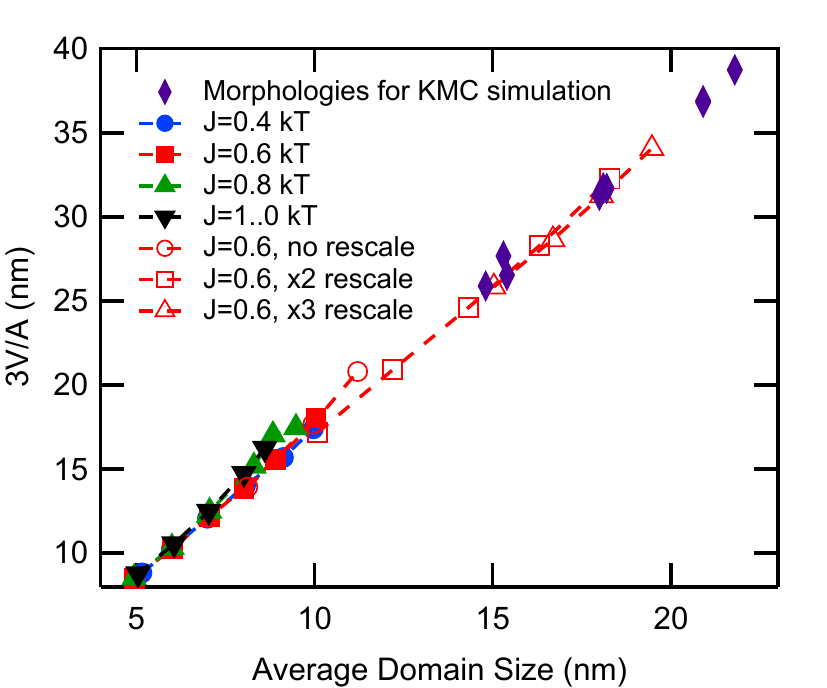}
\caption{\label{fig:figS3} Comparison of domain size calculation methods.  The average domain size determined by the pair-pair correlation method vs.\ the domain size estimated from the average interfacial area to volume ratio.}
\end{figure}

In numerous previous studies, the domain size has been estimated using the relationship, $d=\frac{3V}{A}$, where $V$ is the volume and $A$ is the interfacial area.\cite{marsh2007,groves2008a,groves2008b,groves2009,groves2010a,meng2010,feron2012a,maqsood2013}  This relationship is only strictly valid when the domains are spherical, and since the Ising model produces highly non-spherical domains, this approximation may overestimate the domain size.  For the wide range of morphology sets generated for this study, covering different interaction energies, rescaling factors, and lattice sizes, we have compared the domain size determined by the pair-pair correlation method with that determined from the interfacial area to volume ratio method.  Figure~\ref{fig:figS3} shows that these two techniques have major differences, with the interfacial area to volume ratio method producing a value that is about 75\% greater than the pair-pair correlation method. 

\section{Calculation Time Benchmarks}

To further characterize the improvements in calculation time gained by the various techniques discussed in the main text, we have recorded the calculation time for all morphology sets generated and report several of them here as benchmarks.  All morphologies were generated using the bond formation algorithm that was discussed in the previous section.  For these benchmarks, it is also important to note that the domain size calculation can also take significant calculation time, especially when the domains become large.  To reduce the calculation time required for the domain size characterization, a cutoff distance is used so that the pair-pair correlation function is only calculated out to the cutoff distance.  The minimum calculation time is obtained when the cutoff distance is only slightly larger than the domain size.  For the calculation time benchmark tests, the domain size calculation time was minimized by predicting the approximate domain size that would be obtained for each set of input parameters and using the predicted value to set the initial cutoff distance.  However, if the predicted cutoff was too small, the cutoff distance was incremented and the domain size calculation was performed again.  This same predicted relationship was also used to determine how many MC steps to run the phase separation simulation for to obtain a specified target domain size.

\begin{figure}[h]
\includegraphics[]{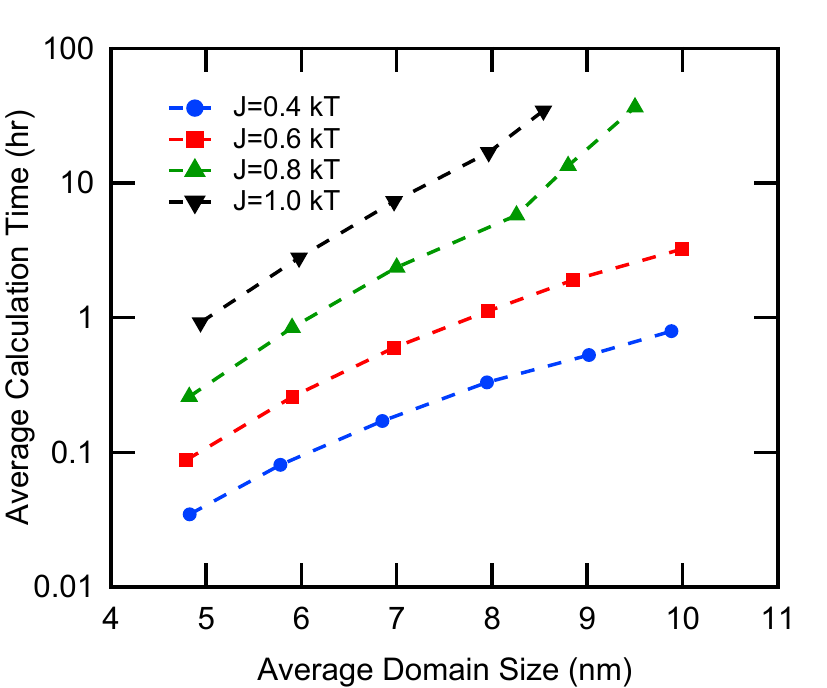}
\caption{\label{fig:figS4} The effect of the interaction energy, $J$, on the calculation time for $J=0.4~kT$ (blue circles), $J=0.6~kT$ (red squares), $J=0.8~kT$ (green triangles), and $J=1.0~kT$ (black inverted triangles).}
\end{figure}

As discussed in the main text, the interaction energy ($J$) significantly changes the rate of domain growth during the simulated phase separation.  As a result of this, generating a morphology of a specific domain size takes considerably different amounts of calculation time depending on the interaction energy used.  Calculation time benchmarks were taken from the morphology sets generated on a 50 by 50 by 50 lattice to determine the average calculation time.  Figure~\ref{fig:figS4} shows how the average calculation time depends on both the desired domain size and the interaction energy.  We find that reducing the interaction energy from $1.0~kT$ to $0.6~kT$ reduces the calculation time by about one order of magnitude for most domain sizes in this range.  We did not test larger lattices here specifically, but the calculation time for all interaction energies should increase linearly with the lattice volume.  

\begin{figure}[h]
\includegraphics[]{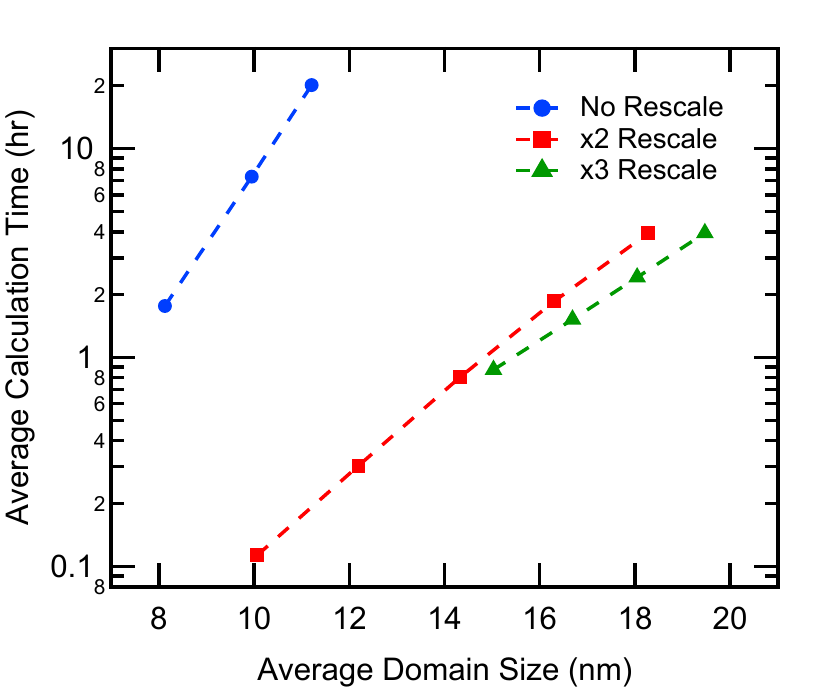}
\caption{\label{fig:figS5} The effect of lattice rescaling on the calculation time without rescaling (blue circles), for x2 rescaling (red squares), and for x3 rescaling (green triangles) created with $J=0.6~kT$.}
\end{figure}

The other major calculation time reductions come into play when using smaller lateral lattice dimensions and the lattice rescaling technique.  When creating large domains on larger lattices, these techniques provide major calculation time reductions.  To test the impact of these techniques, morphology sets of varying domain sizes were created on final lattices with approximately 100~nm height for various rescaling factors.  Lateral lattice dimensions (L) were set to 4.5 times the target domain size rounded up to nearest integer.  For morphology sets without rescaling, L by L by 100 nm lattices were used.  For morphology sets using a rescaling factor of 2, L/2 by L/2 by 50 nm initial lattices were used to reach a final lattice size of L by L by 100 nm.  For a rescaling factor of 3, L/3 by L/3 by 33 initial lattices were used to reach a final lattice size of L by L by 99 nm.  24 morphologies were created under each set of conditions, and the average calculation time was determined.  Figure~\ref{fig:figS5} shows that going from no rescaling to x2 rescaling reduces the calculation time by almost 2 orders of magnitude for generating 10~nm domains.  For larger domain sizes, the x3 rescaling continues to decrease the calculation time compared to x2 rescaling, but the effect is much weaker.

\section{Analysis of Model Bulk Heterojunction Morphologies used for KMC Simulations}

\begin{figure}[h]
\includegraphics[]{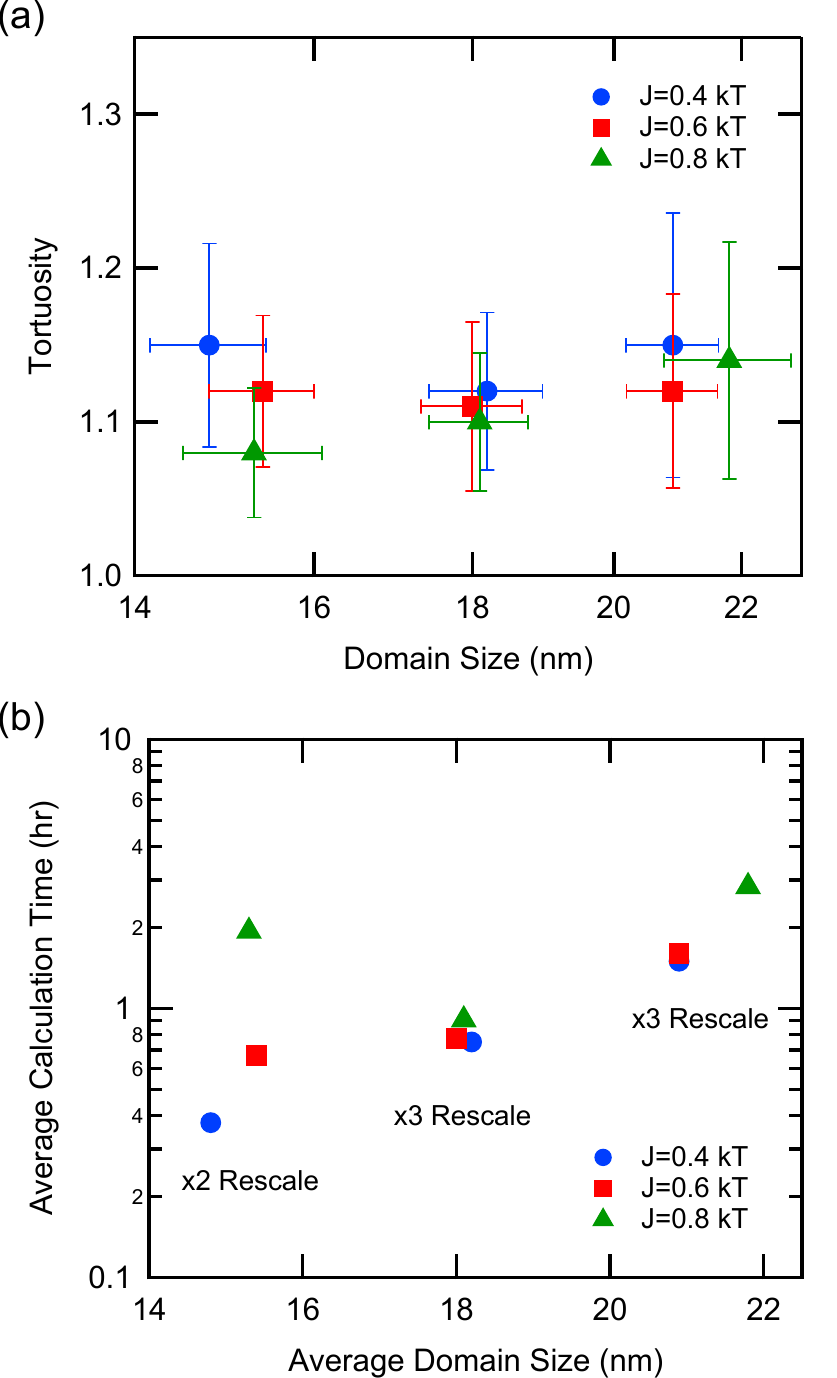}
\caption{\label{fig:figS6}Analysis of morphology sets generated for the KMC benchmark simulations. (a) Tortuosity and (b) average calculation time of morphologies with domain sizes of approximately 15, 18, and 21~nm for $J=0.4~kT$ (blue circles), $J=0.6~kT$ (red squares), and $J=0.8~kT$ (green triangles).  The rescaling factor used for each domain size also indicated.}
\end{figure}

Here, we provide additional characterization of the morphology sets generated for the KMC simulations.  To efficiently create morphologies with 15~nm domains, a 34 by 34 by 51~nm initial lattice was used and rescaled by a factor of 2.  For the morphologies with 18~nm domains, a 27 by 27 by 34~nm initial lattice was used and rescaled by a factor of 3, and for 21~nm domains, a 32 by 32 by 34~nm initial lattice was also rescaled by a factor of 3.  Figure~\ref{fig:figS6}a shows the tortuosity of each morphology set.  As shown for other morphologies in the main text, we find that the tortuosity in general increases as the interaction energy decreases.  However, we also note that for both $J=0.4~kT$ and $J=0.8~kT$, the tortuosity changes with increasing domain size, while for $J=0.6~kT$, the tortuosity is essentially constant.  We find this to be an ideal quality for creating model morphologies because it allows control over the domain size independent from the tortuosity.

Figure~\ref{fig:figS6}b shows the average calculation time required to generate the morphologies.  In general, a lower interaction energy results in a shorter calculation time, but the differences are fairly minor for larger domain sizes.  In addition, going from a rescaling factor of 2 to 3 does not result in a reduced calculation time in these cases.  This is because for the large domain sizes generated here, the domain size calculation step using the pair-pair correlation method is the slowest step.  This limits the calculation time reductions when creating very large domains.  As a result, there is less calculation time difference between different interaction energies in this regime when using rescaling.  In general, though, this shows that creating a typical model morphology for KMC simulations takes about 0.5 to 2 hours per morphology per processor. 

\section{Computational Details}

All computational algorithms were optimized and implemented using C++ and compiled using GCC with speed optimization compiler options.  All calculation time benchmarks were performed on HP Proliant DL 165 G6 computing nodes with two Six-Core AMD Opteron(tm) 2423 HE processors and 16~GB of RAM.  This software has been developed for use with super computing clusters using MPI and can subsequently be used to generate large morphology data sets.  This software is now available open-source on Github,\cite{heiber2014a} and in addition, a user-friendly, web-based morphology generation tool is also available on Nanohub.\cite{heiber2014b}  



\bibliography{references}
